# Impact of gate leakage considerations in tunnel field effect transistor design

Poornendu Chaturvedi[1,2] and M. Jagadesh Kumar[2*]

[1]Nanotechnology Group, Solid State Physics Laboratory, Delhi 110054, India

[2]Indian Institute of Technology, New Delhi 110 016, India

E-mail: poornendu@gmail.com mamidala@ee.iitd.ac.in

## Abstract

In this paper, we have presented the impact of the gate leakage through thin gate dielectrics (SiO$_2$ and high-$\kappa$ gate dielectric) on the subthreshold characteristics of the tunnel field effect transistors (TFET) for a low operating voltage of 0.5 V. Using calibrated two-dimensional simulations it is shown that even for such a low operating voltage, the gate leakage substantially degrades several subthreshold parameters of the TFET such as the off-state current, minimum subthreshold swing and average subthreshold swing. While the drain-offset as well as the short-gate are effective methods for reducing the gate leakage, we show that if the gate tunneling leakage is not considered, even for these two methods, the overall TFET off-state current will be significantly underestimated. Our results demonstrate the need to carefully account for the gate leakage in the design of TFETs just as it is done for the conventional nanoscale MOSFETs.

# 1. Introduction

Devices operated in subthreshold or near-threshold region,[1, 2] are being aggressively explored for reducing both the operating and the stand-by power consumption. Subthreshold design also enables realization of self-powered applications dependent solely on energy scavenging techniques. Tunnel field effect transistors (TFET) are attractive candidates for subthreshold or near-threshold design due to their steep sub-threshold characteristics.[3-14] Leakage currents play a crucial role in this region of operation. Hence, subthreshold and near-threshold circuit design requires much more accurate modelling of leakage currents, particularly in the subthreshold region.

Gate leakage is known to be an important contributor of the off-state leakage in the conventional thin $SiO_2$ gate based MOSFETs,[15, 16]. For the fully depleted silicon-on-insulator (FD-SOI) structures, the ITRS (2012) update predicts an equivalent oxide thickness (EOT) of 0.8 nm going into production for high performance logic in 2014 and for low standby power applications in 2017.[17] Such extremely small EOT will require us to look at gate leakage afresh even for the MOSFETs. Even in a TFET, the gate leakage could be the most important contributor of off-state leakage. Neglecting the gate leakage results in a very small subthreshold slope and gives surprisingly low $I_{OFF}$ and subthreshold slope values which have not been achieved experimentally.[18, 19] High-κ dielectrics with larger dielectric thickness or gate stacks have been used in the TFET to reduce the gate leakage.[20, 21] However, even with the high-κ dielectrics, the fabricated Si TFET devices have not achieved the theoretically predicted subthreshold swing values. Most of the theoretical work on TFET neglects gate leakage. The purpose of this paper is to show that how erroneous it is to assume that there is no gate leakage in TFET. Further, a careful study quantifying the effect of gate leakage on TFET characteristics has also not been undertaken.

In this paper, we have studied the impact of gate leakage on the subthreshold characteristics of TFET with an effective gate dielectric thickness as low as 0.8 nm. For a quantitative validity of our results, we have calibrated both the TFET tunnelling current and gate leakage current with the previous published results. The results in this work show that the gate leakage can substantially alter the subthreshold characteristics of the TFET with both $SiO_2$ and high-κ gate dielectric. We have further studied the effect of gate drain alignment on the TFET off-state current and the gate leakage. TFETs with a drain-offset have been fabricated experimentally for reducing the off-state current[22]. Such a structure also reduces the parasitic capacitances[23, 24]. Another similar method that can be employed for reducing the gate leakage is the short-gate structure[25]. In this work, we show that even for these two methods, the overall TFET off-state current will be underestimated if the gate tunneling leakage is not considered. The details of the device structure and simulation parameters are listed in Sect. 2. Mechanism of the gate leakage and calibration of results are discussed in Sects. 3 and 4, respectively. Effect of the gate leakage on device characteristics is presented in Sect. 5. Finally, Sect. 6 draws important conclusions of this study.

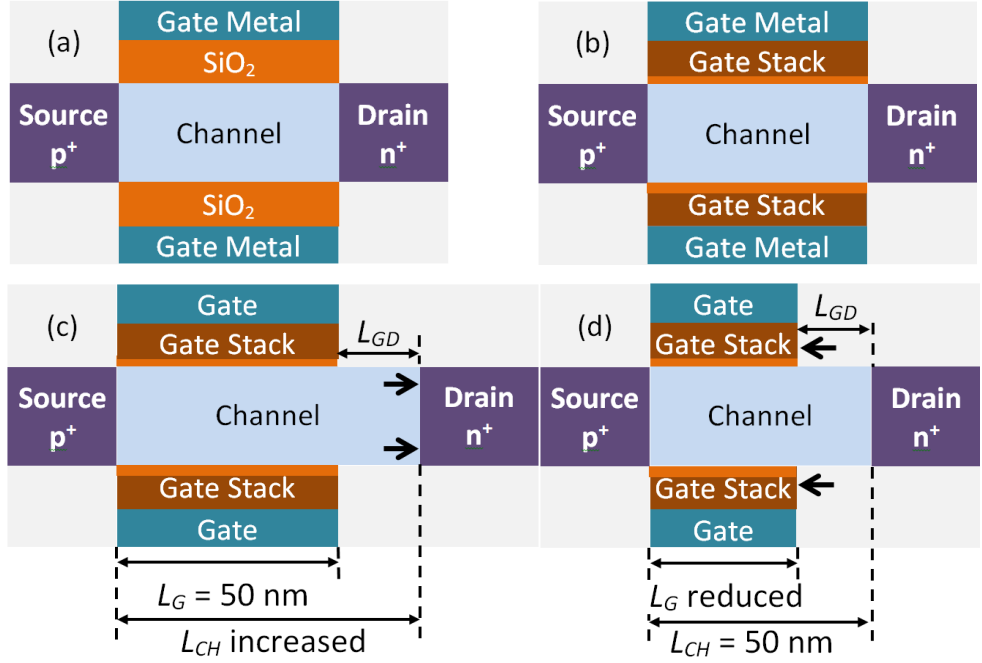

Figure 1: (Color online) Double gate TFET structure with p$^+$ source and n$^+$ drain (a) with SiO$_2$ as gate oxide, (b) with high-κ HfO$_2$/SiO$_2$ gate stack, (c) high-κ gate stack with drain-offset, and (d) high-κ gate stack with short-gate. The parameters $L_G$, $L_{CH}$, and $L_{GD}$ refer to the gate length, the channel length and the gap between the gate edge and the drain edge, respectively. For the structures (a) and (b), $L_{GD} = 0$ and $L_G = L_{CH} = 50$ nm. In (c) and (d), the arrows indicate the direction in which the length of the channel (gate) is increased (decreased).

## 2. Device structure and simulation parameters

Figure 1 shows the structures of the simulated double gate TFET including the drain-offset and the short-gate structures. Both SiO$_2$ [Fig. 1(a)] as well as high-κ HfO$_2$/SiO$_2$ gate stack [Figs. 1(b)-1(d)] were used as the gate dielectric. To include the effect of fringing field in the simulations, gate electrodes of finite thickness, surrounded by the oxide were used. The device parameters are listed in Table I and are based on previous work on TFET.[20] The drain bias ($V_{DS}$) and the gate metal work function ($\Phi_M$) were chosen as 0.5 V and 4.3 eV, unless otherwise stated. Simulations were carried out using a two-dimensional (2D) device simulator.[26] Non-local model for band to band tunnelling (BTBT) was used in the simulations to take into account spatial profile of the energy bands, as well as, spatial separation of the electrons generated in the conduction band from the holes generated in the valence band. Due to the high doping in the source and the drain regions, the effect of band gap narrowing was also included. Fermi Dirac carrier statistics and concentration dependent Shockley-Read-Hall model were used to accurately simulate the device performance. BTBT Tunneling was considered both at the source-channel junction (region of the gate-source overlap) as well as at the drain-channel junction (region of the gate-drain overlap). The tunneling at the drain-channel junction was incorporated to include the impact of gate induced drain leakage on TFET characteristics. Abrupt source channel junction and drain-channel junctions were assumed in all simulations, in line with several other simulation studies.

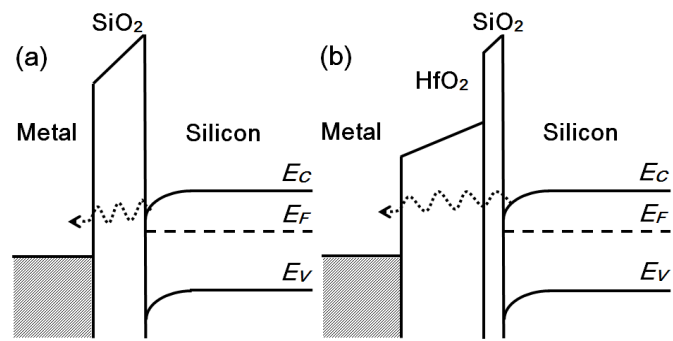

Figure 2: Gate leakage in the TFET through (a) $SiO_2$ and (b) $HfO_2/SiO_2$ gate stack.

## 3. Gate leakage

For SOI structures, the gate leakage can be conveniently divided into tunnelling from the gate into the source and the drain. At low gate voltages, only direct tunnelling of electrons across the gate dielectric takes place from the gate into the drain. As gate voltage increases, tunnelling begins from the source into the gate, reversing the direction of gate current. This also reduces the availability of carriers in the source for $I_{DS}$. Transverse band profiles at the two junctions are also drastically different due to different doping concentrations in the source and drain regions, which affect the quantum of tunnelling in the two regions. This leads to a different slope of $I_G$-$V_{GS}$ curve for the gate-drain tunnelling and the source-gate tunnelling. The leakage through the gate dielectric due to the tunnelling of electrons near the source is illustrated in Fig. 2 for both the $SiO_2$ as well as the high-κ $HfO_2/SiO_2$ gate stack. As the gate voltage is increased, more electrons from the conduction band of Si tunnel through the dielectric. The situation is reversed near the drain due to the applied drain bias, where, electrons tunnel from metal gate into the channel through the dielectric. The tunnelling current density J through the dielectric can be given by[27]

$$J = \frac{qkT}{2\pi^2\hbar^3} \sqrt{m_y m_z} \int T(E) \ln \left\{ \frac{1 + \exp[(E_{Fr} - E)/kT]}{1 + \exp[(E_{Fl} - E)/kT]} \right\} dE, \qquad (1)$$

where q is the charge of the carrier, k is the Boltzmann constant, T is the temperature in Kelvin, h is the Planck's constant, E the energy of the carrier, $m_y$ and $m_z$ are the effective masses in the lateral direction in the semiconductor, $E_{fl}$ and $E_{fr}$ are the quasi-Fermi levels on either side of the barrier, and T(E) is the transmission probability of an electron or hole through the dielectric potential barrier. Schenk oxide tunnelling model was used to calculate the transmission probability or tunnelling through the trapezoidal dielectric barrier.[27] This model also takes into account the image barrier lowering. Extensive calibration of Schenk's oxide tunnelling model as well as BTBT model was carried out with previously published simulated and experimental results.

## 4. TFET calibration

Accurate calibration of the BTBT with experimentally fabricated TFET is a challenging task, as data related to wide design space of TFETs is not easily available for experimental results. Therefore we have calibrated the BTBT model for TFET tunnelling current with Boucart's earlier work,[28] who had calibrated her work with experimentally fabricated IBM tunnel diodes. The calibration was performed simultaneously for κ = 3.9 and 21, corresponding to SiO₂ and HfO₂. The extracted parameters were further used in calibration of gate leakage through both SiO₂ as well as high-κ gate stacks of different thicknesses. For accuracy of gate leakage results, TFET gate leakage through both SiO₂ and HfO₂/SiO₂ gate stacks was calibrated. The gate tunnelling was calibrated with standard MOS capacitor. Even though our device is TFET, the tunnelling through gate oxide will be similar to MOS capacitor as gate tunnelling depends on vertical electric field. The details of the calibrated dielectrics are listed in Table II. We first calibrated the Schenk's oxide tunnelling model with the experimental results for 1.5 nm thick SiO₂ (device 2).[29] To consider more technologically relevant EOT, calibration of Schenk's model with two different high-κ HfO₂/SiO₂ gate stacks with an EOT of 1 nm (devices 3 and 4), was carried out.[30, 31] As mentioned earlier, 0.8 nm EOT gate dielectrics are soon expected to enter production for FD-SOI. It is, therefore, important to study the impact of leakage through gate dielectric with 0.8 nm EOT on the performance of TFET. To take care of this, we have simulated TFET with EOT of 0.8 nm (device 5). To have an acceptable value of physical dielectric thickness, an interfacial SiO₂ layer of 0.4 nm and HfO₂ layer of 2.0 nm were used. The calibrated gate leakage for an EOT of 0.8 nm was found to be of the same order of magnitude as the previously published experimental result with similar EOT.[32] Comparison of the calibrated gate leakage with the previous published works is shown in Fig. 3. Excellent match of calibrated gate leakage with previous published works was achieved before undertaking the current study. We have been able to calibrate Schenk's oxide tunnelling model over more than eight orders of magnitude of gate current density from $\sim 10^{-13}$ to $10^{-5}$ A/$\mu m^2$, for different dielectric structures, both single layer $SiO_2$ and multilayer $HfO_2/SiO_2$ high-$\kappa$ gate stacks.

Although, in the past the Schenk's model has been used for single layer oxides, after calibration it was found to accurately predict the gate leakage of even the double layer $HfO_2/SiO_2$ gate stacks, at least for the region of TFET operation used in this study. For comparison with high-$\kappa$ gate stacks, leakage through $SiO_2$ of 1 nm (Device 1) was also calibrated.[30] As the leakage through both 1.0 and 1.5 nm $SiO_2$ could be calibrated with similar tunnelling parameters, these parameters are also expected to be valid for the intermediate dielectric thicknesses. The details of calibration and calibrated parameters are listed in Appendix.

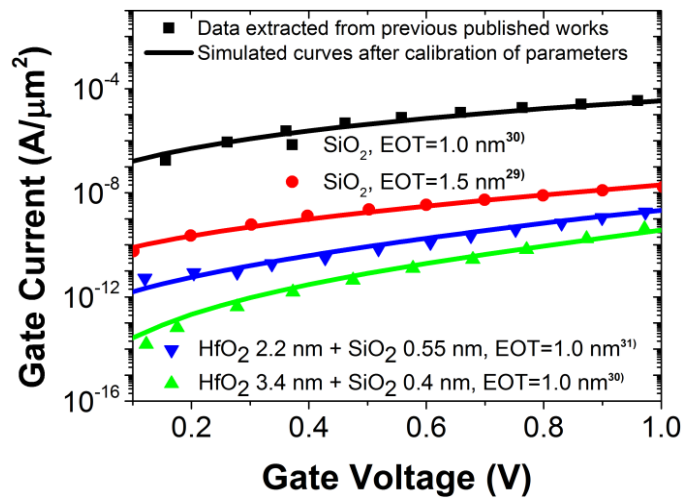

Figure 3: (Color online) Calibration of the gate leakage for different dielectrics with previous published works.

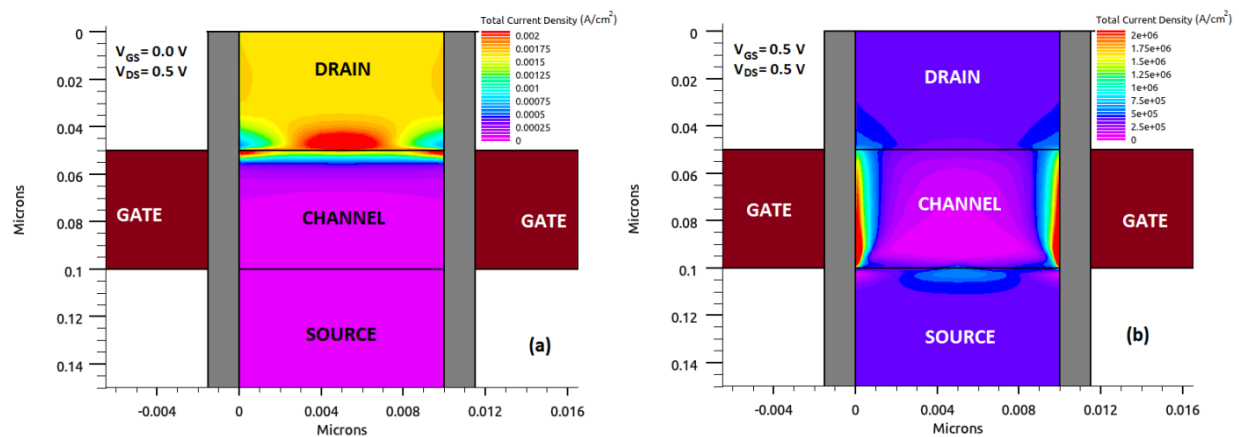

Figure 4: (Color online) Spatial distribution of the total current density in the TFET (with gate oxide thickness of 1.5 nm) in (a) the off-state and (b) the on-state.

## 5. Results and discussion

Figure 4 shows the spatial distribution of the total current density in the TFET at two different bias voltages. In the off-state ($V_{GS}$ = 0.0 V and $V_{DS}$ = 0.5 V), (i) the gate leakage is the dominant component and appears in the vicinity of the drain and (ii) there is no significant tunneling current from the source to the channel or gate induced drain leakage from the drain into the channel. In the on-state ($V_{GS}$ = 0.5 V and $V_{DS}$ = 0.5 V), (i) the tunneling current from the source is the dominant component and appears in the vicinity of the source and (ii) the gate leakage is not apparent as it is overshadowed by the TFET tunneling current that is several orders of magnitude larger.

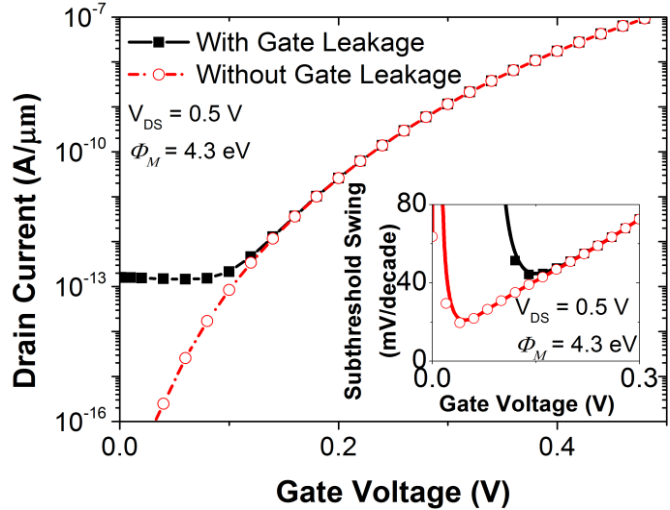

Figure 5: (Color online) Impact of the gate leakage on the $I_D$-$V_{GS}$ characteristics and the subthreshold swing of the TFET.

The $I_D$-$V_{GS}$ characteristics of the TFET for the SiO$_2$ thickness of 1.5 nm and the $\Phi_M$ of 4.3 eV are shown in Fig. 5. At low gate bias voltages, gate leakage is the dominating component of drain current until superseded by BTBT from source into the channel. The gate leakage not only dominates the other leakage currents, but it is several orders of magnitude larger than even the initial tunnelling current at the drain-channel junction. Therefore, the gate to drain tunnelling determines the effective $I_{OFF}$ of the TFET. If the gate leakage is not included, the off-state current can be attributed to the tunnelling at the drain channel junction near the gate. The $I_D$-$V_{GS}$ characteristics of the TFET show $I_{OFF} < 10^{-16}$ A/μm. With inclusion of gate leakage, $I_{OFF}$ increases to ~ $10^{-13}$ A/μm, showing that the gate leakage is more than three orders of magnitude larger than the other leakage currents. As the gate voltage is increased further, tunnelling from the source into the gate increases. This gate tunnelling, which occurs at the gate edge near the source, determines the gate leakage under ON conditions. This leakage is considerably small compared to the device ON current for SiO$_2$ thickness of 1 nm and above. This is reflected by the negligible effect of gate leakage on device ON current. The subthreshold swing (SS) for TFET defined as

$$SS = \frac{dV_{GS}}{d(\log I_D)} \text{ mV/decade,} \qquad (2)$$

which is also shown in the inset of Fig. 5, for the cases with and without considering the gate leakage. The subthreshold swing minimum (SS$_{MIN}$) occurs when the bands are just getting aligned and the tunnelling probability increases rapidly. Due to the exponential dependence of tunnelling on the band alignment, TFET have a sharp turn on characteristics. With the inclusion of the gate leakage in the device simulations, a significant portion of the sharp turn-on of TFET is lost. When the gate leakage is not considered, the inset of Fig. 5 shows the SS$_{MIN}$ to be 20 mV/decade. However, by including the gate leakage, SS$_{MIN}$ more than doubles to 45 mV/decade.

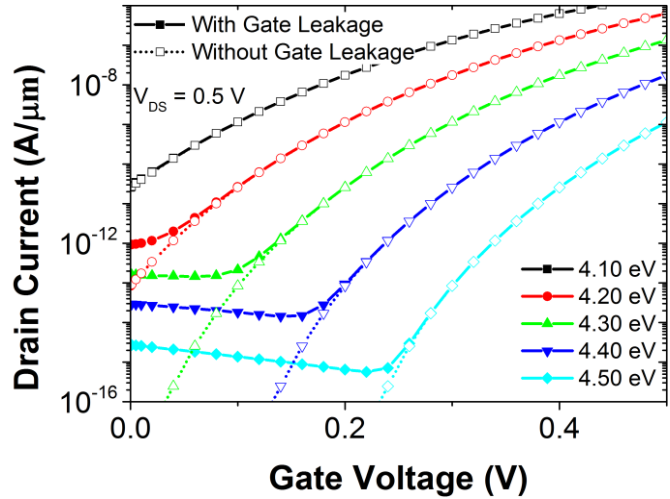

Figure 6: (Color online) Impact of the gate leakage on the TFET with different gate metal work functions and for SiO$_2$ thickness of 1.5 nm.

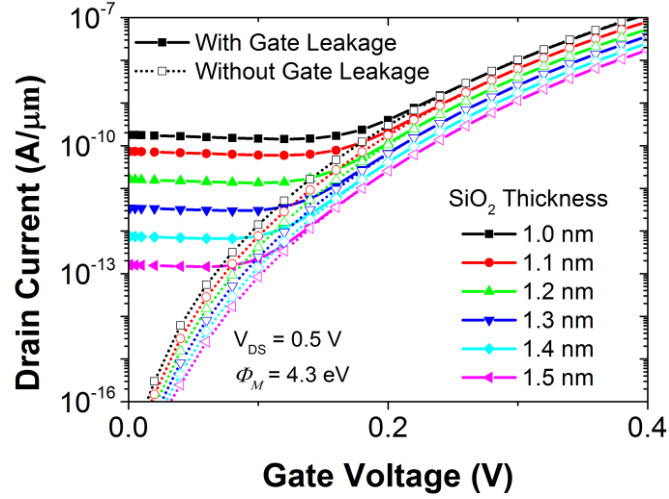

Figure 7: (Color online) Effect of SiO$_2$ thickness on the I$_D$-V$_{GS}$ characteristics, both with and without including the gate leakage.

The parameter that is more important than the point subthreshold swing value of SS$_{MIN}$ is the average subthreshold swing (SS$_{AVG}$), defined over several decades of the drain current. We propose that the point of SS$_{MIN}$ can be taken as the point at which the TFET turns on. This also allows to better define SS$_{AVG}$. The average subthreshold swing can be defined between the point the TFET turns on and the point it is able to deliver a drain current of 10$^{-7}$ A/μm. Mathematically SS$_{AVG}$ can be defined as

$$SS_{AVG} = 1000 * \frac{V_G@(I_D=10^{-7}) - V_G@ss_{MIN}}{\log(10^{-7}) - \log I_D(SS_{MIN})} \, mV/decade, \quad (3)$$

where I$_D$(SS$_{MIN}$) is the drain current when subthreshold slope is minimum and V$_G$@SS$_{MIN}$ is the gate voltage at this point. After including the gate leakage, SS$_{AVG}$ increases by more than 38% from 52 mV/decade over seven orders of magnitude to 72 mV/decade over just four orders of magnitude. Gate leakage has also shifted the point of minimum subthreshold swing (V$_G$@SS$_{MIN}$) by more than 100 mV, necessitating extra 0.1 V to start tunnelling. These observations are extremely important for subthreshold circuit design. Thus, gate leakage severely restricts the advantages of TFET due to the higher operating voltage and higher leakage current. The impact of the gate metal work function on the gate leakage and the TFET characteristics is shown in Fig. 6. After inclusion of the gate leakage, significant differences emerge in the subthreshold characteristics of TFET for Φ$_M$ > 4.1 eV. Gate metal work function has a significant impact on the TFET characteristics. The I$_{ON}$ and I$_{OFF}$ can be adjusted by altering the gate metal work function. Reducing the gate work function from 4.5 to 4.2 eV enhances both I$_{ON}$ and I$_{OFF}$ by two orders of magnitude.

The thickness of the $SiO_2$ was varied (from 1.0 to 1.5 nm in steps of 0.1 nm) to understand the relationship between the dielectric thickness and the gate leakage. The effect of $SiO_2$ thickness ($t_{ox}$) on TFET characteristics is depicted in Fig. 7. As expected, both $I_{ON}$ and $I_{OFF}$ are strong functions of the $SiO_2$ thickness. This is due to the exponential dependence of both BTBT tunnelling and oxide tunnelling on the electric field. It can also be seen that $I_{OFF}$ increases more rapidly than $I_{ON}$ with a reduction in $t_{ox}$. For a 1 nm thick $SiO_2$, gate leakage causes $I_{OFF}$ to increase by more than six orders of magnitude.

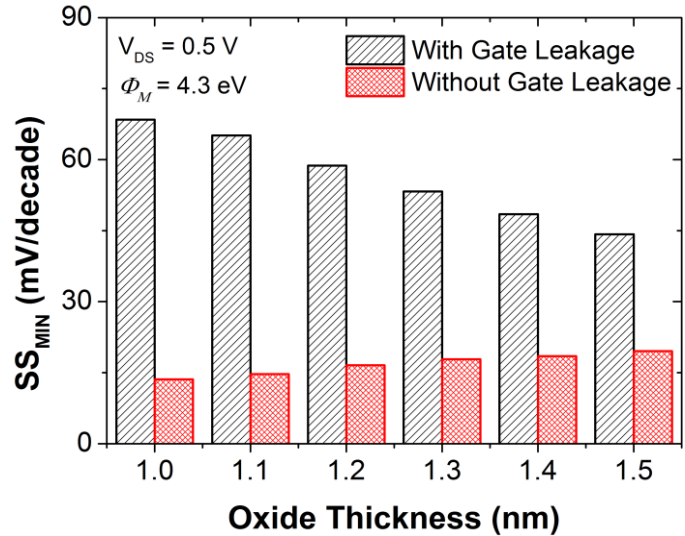

Figure 8: (Color online) $SS_{MIN}$ for different $SiO_2$ thickness, both after neglecting and including the gate leakage.

Figure 8 shows the values of $SS_{MIN}$ for different $SiO_2$ thicknesses. The impact of gate leakage on the subthreshold characteristics increases with reducing $t_{ox}$. An interesting observation is the relative order of subthreshold swing minimum ($SS_{MIN}$) for different $SiO_2$ thicknesses. Without considering the gate leakage, $SS_{MIN}$ increases by increasing the $SiO_2$ thickness. However, when the gate leakage is included, TFET with thicker $SiO_2$ has a lower value of $SS_{MIN}$, indicating a sharp turn-on. This is caused by the reduced gate leakage at a higher $SiO_2$ thickness. This allows BTBT current to supersede the gate leakage at lower gate voltages. Hence, devices with lower $SiO_2$ thickness have higher $I_{ON}$, but their subthreshold characteristics are compromised.

Current devices employ high-$\kappa$ gate stacks with $HfO_2$ as gate dielectric and $SiO_2$ as interfacial layer for reducing gate leakage by having larger physical thickness and lower EOT. We have also studied the impact of gate leakage on the TFET with different high-$\kappa$ gate stacks having an EOT of 0.8 nm to 1.0 nm, as mentioned earlier in Table 2. The EOT of the $HfO_2/SiO_2$ gate stack is given by

$$\text{EOT} = \left( t_{\text{HfO}_2} \cdot \frac{k_{\text{SiO}_2}}{k_{\text{HfO}_2}} + t_{\text{SiO}_2} \right), \qquad (4)$$

where t and κ refer to the thickness and the dielectric constant of HfO₂ and SiO₂, respectively. The I$_D$-V$_{GS}$ characteristics for the devices 3 - 5 are shown in Fig. 9, both with and without including the gate leakage. It can be seen that although the gate leakage is much lower in HfO₂ stacks compared to SiO₂, still incorporation of gate leakage in TCAD simulations significantly alters the TFET characteristics.

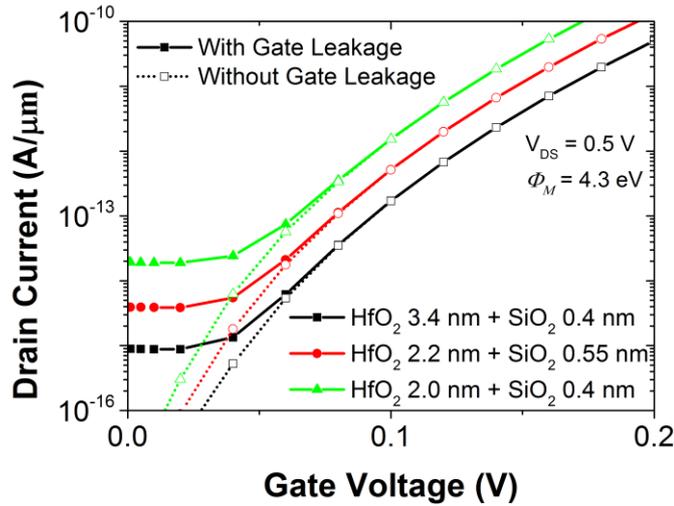

Figure 9: (Color online) I$_D$ − V$_{GS}$ characteristics of the TFET with a high-κ gate stack.

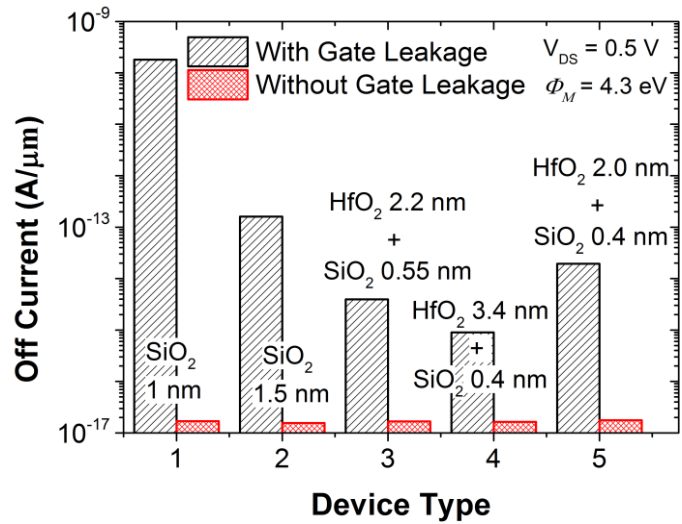

Figure 10: (Color online) Impact of the gate leakage on the I$_{OFF}$ of the TFET for different device types.

The effect of gate leakage on I$_{OFF}$ is shown in Fig. 10 for all the five devices. Even for device 3 having an EOT of 1.0 nm with 2.2 nm HfO₂ and 0.55 nm SiO₂, I$_{OFF}$ increases by more than two orders of magnitude, if gate leakage is included. From Eq. (4), we can also see that by reducing the thickness of the interfacial layer, the physical thickness of HfO$_2$ stack can be increased for the same EOT. Device 4 makes use of this phenomenon and has a total gate dielectric thickness of 3.8 nm and an EOT of 1 nm. This causes the gate leakage to reduce further. However, even with such thick gate dielectric, gate leakage increases I$_{OFF}$ by more than an order of magnitude. Even this advantage is compromised when the EOT was reduced to 0.8 nm and I$_{OFF}$ was found to increase by more than three orders of magnitude.

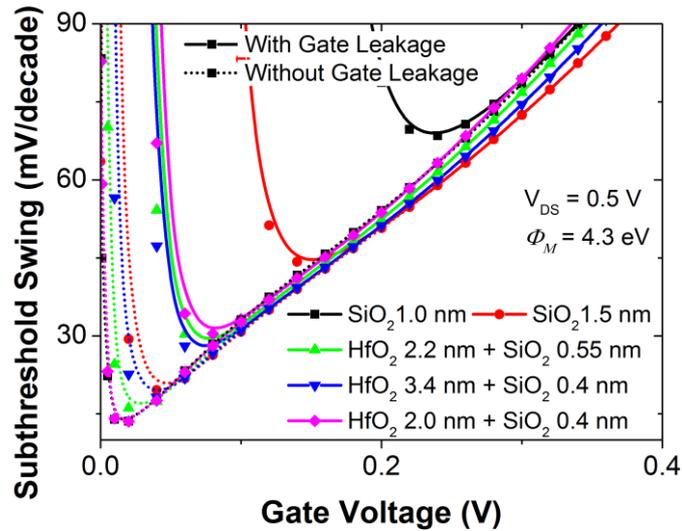

Figure 11: (Color online) Effect of the gate leakage on the subthreshold swing of the TFET for different gate dielectrics.

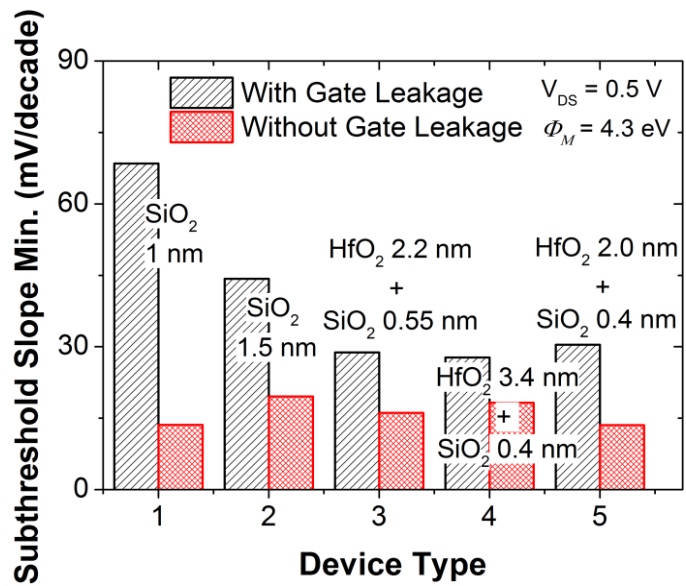

Figure 12: (Color online) Impact of the gate leakage on the SS$_{MIN}$ for different device types.

Figures 11 and 12 show the impact of the gate leakage on the subthreshold characteristics for the five device types. For devices 1, 3, and 4, having an EOT of 1 nm, the inclusion of gate leakage in TFET simulations increases $SS_{MIN}$ by 402, 78 and 52% and increases $SS_{AVG}$ by 90, 29 and 16%, respectively. Once again it can be observed that even with high-$\kappa$ gate stacks with an EOT of 1 nm, gate leakage not only affects the $I_{OFF}$, but it also increases the $SS_{MIN}$, $SS_{AVG}$ and shifts the gate voltage at which device turns on ($V_G@SS_{MIN}$). For device 5, having EOT of 0.8 nm, $SS_{AVG}$ also increases by 31%, while $SS_{MIN}$ increases by 124%, from 13.6 to 30.4 mV/decade, showing a substantial degradation in TFET subthreshold performance parameters.

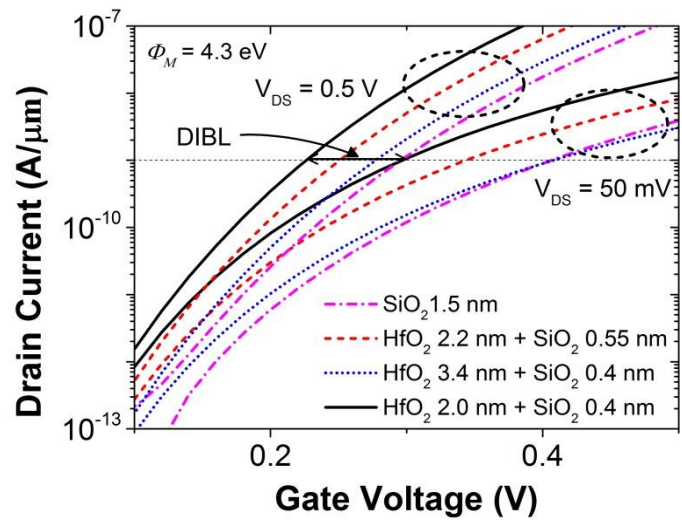

Figure 13: (Color online) Effect of drain bias on the $I_D$-$V_{GS}$ characteristics for different gate dielectrics.

Figure 13 shows the impact of drain bias on $I_D$-$V_{GS}$ characteristics of TFET having high-$\kappa$ gate stacks with an EOT of 1.0 and 1.5 nm thick $SiO_2$. Application of drain bias reduces the $V_{GS}$ required to achieve similar values of drain current, indicating the presence of drain induced barrier lowering (DIBL). The reduction of $V_{GS}$ is 111, 94, 131, and 71 mV in devices 2, 3, 4, and 5 respectively, showing that the impact of DIBL is lowest in device type 5 on account of stronger gate control.

For all the previous simulations the edge of the gate was aligned with the edge of the drain. The gate leakage can be reduced by introducing a gap ($L_{GS}$) between the edge of the gate and the edge of the drain. This can be achieved in two ways, (a) by keeping the gate length ($L_G$) constant and moving the drain away from the gate edge, thereby increasing the channel length ($L_{CH}$), as shown in Fig. 1(c), or (b) by keeping the channel length, $L_{CH,}$ constant and reducing the gate length, $L_G$, as shown in Fig. 1(d). The two approaches are similar, the first approach is called the drain-offset and the second approach is called the short-gate.

The impact of gate leakage on the off-state current, for the drain-offset and the short-gate approaches is shown in Figs. 14(a) and 14(b), respectively. Simulations were carried out both with and without considering gate leakage for device 3 having an EOT of 1 nm. For drain-offset approach, the gap between the gate edge and the drain edge was varied from 0 nm (aligned gate) to 25 nm, keeping the $L_G$ fixed at 50 nm [Fig. 14(a)]. For the short-gate approach, the gate length was varied from 50 nm (aligned gate) to 25 nm, keeping the $L_{CH}$ fixed at 50 nm [Fig. 14(b)].

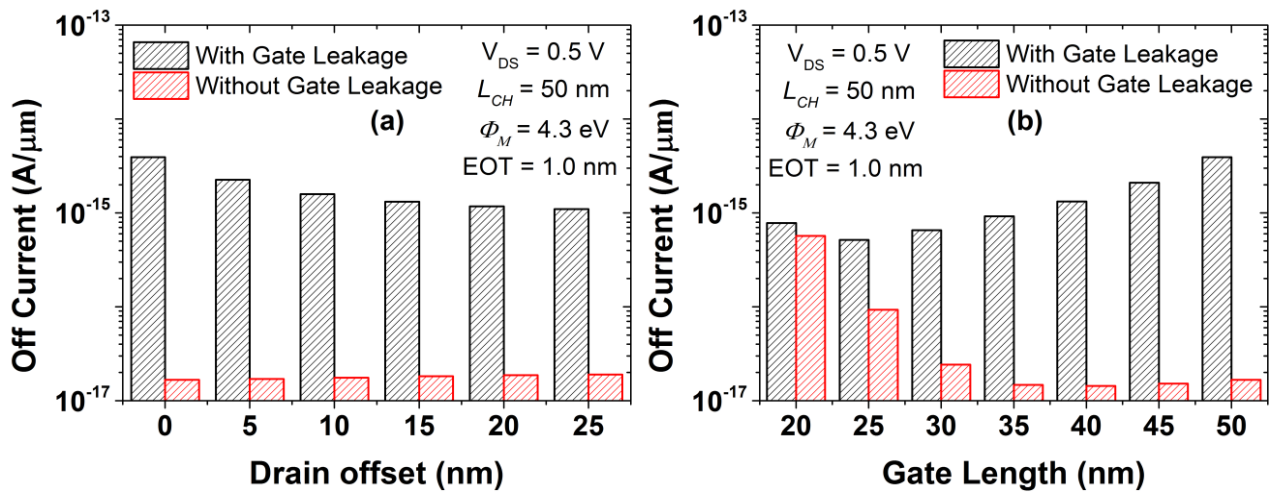

Figure 14: (Color online) Impact of (a) the drain-offset and (b) the short-gate, on the off-state current for device type 3, both with and without considering the gate leakage. The drain offset of 0 nm in (a) and the gate length of 50 nm in (b) refer to the TFET with an aligned gate.

From Fig. 14(a), we can observe that when the gate leakage is not included in the simulations, introducing a drain-offset up to 25 nm has no significant impact on $I_{OFF}$, which remains at ~ 2 x $10^{-17}$ A/μm. After including the gate leakage in the simulations, the drain-offset of 25 nm is successful in reducing $I_{OFF}$ by a factor of 4 in comparison to the aligned gate (drain-offset of 0 nm). However, inclusion of the gate leakage in the simulations increases $I_{OFF}$ by a factor of ~200 for the aligned gate, and by a factor of ~50 for the drain-offset of 25 nm. It may also be noted that increasing the drain-offset beyond 10 nm has a diminishing impact on the off-state current.

In the case of the short-gate TFET [Fig. 14(b)], when the gate leakage in not considered, the off-state current starts to increase as the gate length is reduced below 35 nm. This occurs due to an increase in the BTBT tunneling at the source-channel junction. However, when the gate leakage is included in the simulations, the $I_{OFF}$ decreases as the gate length is reduced from 50 to 25 nm and then increases marginally for the gate length of 20 nm. For $L_G = 20$ nm, the $I_{OFF}$ both with and without including the gate leakage are comparable as the BTBT tunneling is the major contributor of $I_{OFF}$ at this gate length. Compared to the aligned gate ($L_{CH} = 50$ nm and $L_G = 50$ nm), TFET with a 25 nm shorter gate ($L_{CH} = 50$ nm and $L_G = 25$ nm) has almost an order of magnitude smaller off-state current of 5 x $10^{-16}$ A/µm. However, it is still higher by a factor of five compared to the case when the gate leakage is neglected in the simulations. Hence, it can be concluded that for the studied device parameters, the inclusion of the gate leakage current will critically affect the estimation of the off-state current in both the drain-offset approach and the short-gate approach, highlighting once again the importance of including gate leakage particularly for structures with EOT of 1 nm and below.

## 6. Conclusions

In this work, we have shown that the gate leakage severely degrades the TFET subthreshold parameters such as $I_{OFF}$, $SS_{MIN}$, $SS_{AVG}$, and $V_G@SS_{MIN}$. The results were found to be valid for all gate dielectric structures, both $SiO_2$ and high-κ gate stacks, studied in this work. The gate leakage has a substantial effect on TFET with sub 1.5 nm $SiO_2$ gate dielectric. It causes $I_{OFF}$ to increase by three orders of magnitude.

The impact of gate leakage can be reduced with high-κ gate stacks. However, even with 3.8 nm thick gate stack with an EOT of 1 nm, gate leakage increases $I_{OFF}$ by almost two order of magnitude and $SS_{MIN}$ by 52%. For an EOT of 0.8 nm, gate leakage increases $I_{OFF}$ by three orders of magnitude, and $SS_{MIN}$ and $SS_{AVG}$ by 124 and 31%, respectively. It can therefore be concluded that with the use of high-κ stacks the impact of gate leakage can be reduced, but it cannot be eliminated. We have also studied the two commonly employed TFET strategies, namely the drain-offset and the short-gate and found that the inclusion of gate leakage will significantly affect the estimation of $I_{OFF}$, even for these structures with an EOT of 1 nm and below. Moreover, future technology nodes are expected to have much lower EOT than 1 nm, indicating that gate leakage would be an even more important phenomenon for future TFET design, particularly for obtaining their subthreshold characteristics. Hence, it is very important to account for the gate leakage for preventing any unrealistic estimation of TFET parameters. The results also highlight the need for improved TFET designs for minimizing the gate leakage to achieve the true potential of TFET.

Acknowledgements

Appendix

To ensure the validity of the results, extensive calibrations of simulation parameters for BTBT and the gate leakage with previous published works were carried out. First, the BTBT model was simultaneously calibrated for $\kappa$ = 3.9 and 21 with Boucart's earlier work.[28] The extracted tunnelling masses of the electrons [$m_{e,tunnel}$ (Si)] and the holes [($m_{h,tunnel}$ (Si)] are listed in Table A-I, and were used in all further calibrations of gate leakage for all the reference device structures, i.e. device 1,[30] device 2,[29] device 3,[30] and device 4,[31] and TFET simulations.

Table A-I: BTBT model calibration parameters.

| $m_{e,tunnel}$ (Si) | 0.1 | $m_{h,tunnel}$ (Si) | 0.17 |
|---|---|---|---|

For all the above mentioned devices, the gate leakage was calibrated with a MOS structure using parameters listed in Table A-II to Table A-V, respectively. For gate dielectrics, tunnelling masses of the electrons [$m_{e,tunnel}$ ($SiO_2$) and $m_{e,tunnel}$ ($HfO_2$)] and the holes [($m_{h,tunnel}$ ($SiO_2$) and $m_{e,tunnel}$ ($HfO_2$)] were kept equal. The physical parameters of the gate dielectrics were kept similar to the original work. n-type body doping was chosen to improve convergence of simulations.

Table A-II: Device 1 calibration parameters.

| t, $SiO_2$ (nm) | 1.0 | $\chi$, $SiO_2$ (eV) | 0.89 |
|---|---|---|---|
| $m_{e,tunnel}$ ($SiO_2$) | 0.77 | $m_{h,tunnel}$ ($SiO_2$) | 0.77 |
| $\Phi_M$ (eV) | 4.1 | Body doping (n) | $10^{15}$/$cm^3$ |

Table A-III: Device 2 calibration parameters.

| t, $SiO_2$ (nm) | 1.5 | $\chi$, $SiO_2$ (eV) | 0.89 |
|---|---|---|---|
| $m_{e,tunnel}$ ($SiO_2$) | 0.77 | $m_{h,tunnel}$ ($SiO_2$) | 0.77 |
| $\Phi_M$ (eV) | 4.0 | Body doping (n) | $10^{15}$/$cm^3$ |

Tunnelling mass for the electrons ($m_{e,tunnel}$, $HfO_2$), and gate metal work function ($\Phi_M$) were used to calibrate gate leakage. Electron affinity of $SiO_2$ ($\chi$, $SiO_2$) was varied between acceptable limits (0.89 and 1.0) to improve fit with previous published results. Although device 4 corresponds to an EOT of 1.06 nm, value of 1.0 nm was used in this study, as done in the original work.

Table A-IV: Device 3 calibration parameters.

| t, SiO$_2$ (nm) | 0.55 | $\chi$, SiO$_2$ (eV) | 1.0 |
|---|---|---|---|
| t, HfO$_2$ (nm) | 2.2 | $\chi$, HfO$_2$ (eV) | 2.7 |
| m$_{e,tunnel}$ (HfO$_2$) | 0.35 | m$_{h,tunnel}$ (HfO$_2$) | 0.35 |
| $\Phi_M$ (eV) | 4.1 | Body doping (n) | 10$^{17}$/cm$^3$ |

Table A-V: Device 4 calibration parameters.

| t, SiO$_2$ (nm) | 0.4 | $\chi$, SiO$_2$ (eV) | 1.0 |
|---|---|---|---|
| t, HfO$_2$ (nm) | 3.4 | $\chi$, HfO$_2$ (eV) | 2.7 |
| m$_{e,tunnel}$ (HfO$_2$) | 0.2 | m$_{h,tunnel}$ (HfO$_2$) | 0.2 |
| $\Phi_M$ (eV) | 4.3 | Body doping (n) | 10$^{15}$/cm$^3$ |

Since, the Schenk's model does not take into account multilayer dielectrics these tunnelling masses do not represent the true tunnelling masses, but only an empirical fitting parameter. For simulating gate leakage through an EOT of 0.8 nm, tunnelling masses were linearly extrapolated using following equation, obtained from previous two calibrations:

$$m_{HfO_2} = -0.1429 \left( t_{HfO_2} + t_{SiO_2} \right) + 0.7429. \qquad (A.1)$$

Device 5 calibration parameters are listed in Table A-VI. The calibrated gate leakage was found to be of the same order of magnitude agree as the previous published experimental result[32] of similar EOT at V$_G$ = 1.0 V. Still, the nonphysical basis of above fitting parameters and limited range of applicability must be kept in mind, while extending these calibrated parameters to other thicknesses and dielectrics.

Table A-VI: Device 5 calibration parameters.

| t, SiO$_2$ (nm) | 0.4 | $\chi$, SiO$_2$ (eV) | 1.0 |
|---|---|---|---|
| t, HfO$_2$ (nm) | 2.0 | $\chi$, HfO$_2$ (eV) | 2.7 |
| m$_{e,tunnel}$ (HfO$_2$) | 0.4 | m$_{h,tunnel}$ (HfO$_2$) | 0.4 |
| $\Phi_M$ (eV) | 4.3 | Body doping (n) | 10$^{15}$/cm$^3$ |

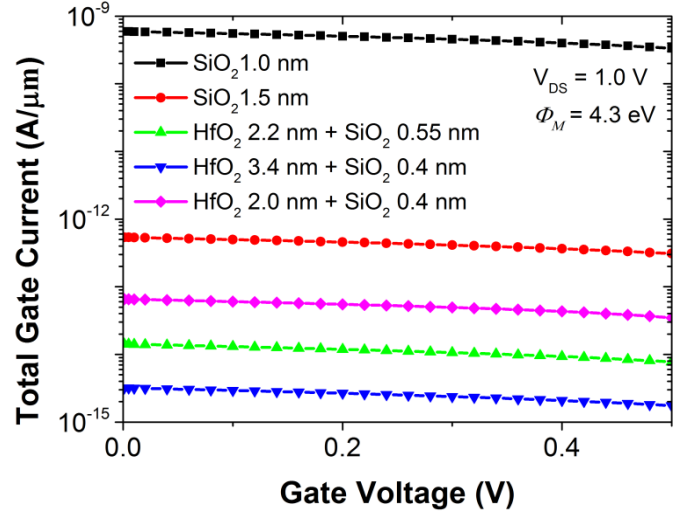

Figure A.1: (Color online) Calibrated $I_G$-$V_G$ characteristics of the TFET with different gate dielectrics.

The $I_G$-$V_{GS}$ characteristics of the TFET for all the five device types are shown in Fig. A.1 depicting gate leakage for five device structures after calibration.

References

[1] D. Bol, R. Ambroise, D. Flandre, and J.-D. Legat, IEEE Trans. VLSI Syst., **17**, 1508 (2009).

[2] H. Kaul, M. Anders, S. Hsu, A. Agarwal, R. Krishnamurthy, and S. Borkar, Proc. 49th Annu. Design Automation Conf., 2012, p. 1153.

[3] L. Wang and P. Asbeck, Proc. Nanotechnology Materials and Devices Conf., 2009, p. 196.

[4] K. K Bhuwalka, J. Schulze, and I. Eisele, IEEE Trans. Electron Devices **52**, 909 (2005)

[5] Y. Khatami and K. Banerjee, IEEE Trans. Electron Devices **56**, 2752 (2009).

[6] A. S. Verhulst, W. G. Vandenberghe, K. Maex, and G. Groeseneken, Appl. Phys. Lett. **91**, 053102 (2007).

[7] A. C. Seabaugh, Z. Qin, Proc. IEEE **98**, 2095 (2010).

[8] S. Saurabh and M. J. Kumar, Jpn. J. Appl. Phys. **48**, 064503 (2009).

[9] S. Saurabh and M. J. Kumar, IEEE Trans. Device Mater. Reliab. **10**, 390 (2010).

[10] S. Saurabh and M. J. Kumar, IEEE Trans. Electron Devices **58**, 404 (2011).

[11] M. C. Sun, S. W. Kim, H. W. Kim, H. Kim and B. G. Park, Jpn. J. Appl. Phys. **52**, 06GE06 (2013).

[12] M. H. Lee, J. C. Lin, C. Y. Kao and C. W. Chen, Jpn. J. Appl. Phys. **52**, 04CC27 (2013).

[13] Y. Morita, T. Mori, S. Migita, W. Mizubayashi, A. Tanabe, K. Fukuda, M. Masahara and H. Ota, Jpn. J. Appl. Phys. **52**, 04CC25 (2013).

[14] M. J. Kumar and S. Janardhanan, IEEE Trans. Electron Devices **60**, 3285 (2013).

[15] K. Roy, S. Mukhopadhyay, H. Mahmoodi-Meimand, Proc. IEEE **91**, 305 (2003).

[16] Y. C. Yeo, T. J. King and C. Hu, IEEE Trans. Electron Devices **50**, 1027 (2003).

[17] ITRS 2012 update (2012).

[18] S. Migita and H. Ota, Proc. Int. Semiconductor Device Research Symp. 2011, p. 1.

[19] M. Fulde, A. Heigl, M. Weis, M. Wirnshofer, K. v. Arnim, Th. Nirschl, M. Sterkel, G. Knoblinger, W. Hansch, G. Wachutka and D. Schmitt-Landsiedel, 2nd IEEE Int. Nanoelectronics Conf., 2008, p. 579.

[20] K. Boucart and A. M. Ionescu, Proc. 36th European Solid-State Device Research Conf., 2006, p. 383.

[21] D. Leonelli, A. Vandooren, R. Rooyackers, A. S. Verhulst, S. De Gendt, M. M. Heyns, and G. Groeseneken, Jpn. J. Appl. Phys. **49**, 04DC10 (2010).

[22] T. Mori, T. Yasuda, T. Maeda, W. Mizubayashi, S. O'uchi, Y. Liu, K. Sakamoto, M. Masahara, and Hi. Ota, Jpn. J. Appl. Phys. **50**, 06GF14 (2011).

[23] Y. Yang, X. Tong, L. Yang, P. Guo, L. Fan, G. S. Samudra, and Y.-C. Yeo, Ext. Abst. Int. Conf. Solid State Devices and Materials, 2009, p. 597.

[24] Y. Yang, X. Tong, L.-T. Yang, P.-F. Guo, L. Fan, and Y.-C. Yeo, IEEE Electron Device Lett. **31**, 752 (2010).

[25] J. Zhuge, A. S. Verhulst, W. G. Vandenberghe, W. Dehaene, R. Huang, Y. Wang, G. Groeseneken, Semicond. Sci. Technol. **26,** 085001 (2011).

[26] ATLAS User's Manual (Silvaco Inc. 2014).

[27] A. Schenk and G. Heiser, J. Appl. Phys. **81**, 7900 (1997).

[28] K. Boucart and A. M. Ionescu, IEEE Trans. Electron Devices **54**, 1725 (2007).

[29] W.-B. Kim, T. Matsumoto and H. Kobayashi, Nanotechnology **21**, 115202, (2010).

[30] F. Li, S. P. Mudanai, Y.-Y. Fan, L. F. Register, and S. K. Banerjee, IEEE Trans. Electron Devices **53**, 1096 (2006).

[31] T. Ando, N. D. Sathaye, K. V. R. M. Murali, and E. A. Cartier, IEEE Electron Device Lett. **32**, 865 (2011).

[32] H. Z. Hong, K. S. C. Liao, C. H. Fu, C. C. Li, Y. Y. Hsu, and T. K. Wang, Proc. Int. Semiconductor Device Research Symp. 2011, p. 1.